\documentclass[10pt]{blois07}
\usepackage{graphicx}
\usepackage{cite,./mcite}
\usepackage{epsfig}

\begin{document}
\title{Ultra-Peripheral Collisions at RHIC}
\author{Joakim Nystrand$^1$\protect\footnote{\hspace{2mm}talk presented at EDS07}}
\institute{$^1$Department of Physics and Technology, University of Bergen, Bergen, Norway}
\maketitle
\begin{abstract}
This presentation summarizes the results on ultra-peripheral collisions 
obtained at RHIC. It also discusses some aspects of the 
corresponding electromagnetic interactions in $pp$  and $p \overline{p}$ 
collisions. 
\end{abstract}

Ultra-peripheral nucleus-nucleus collisions are defined as 
collisions in which the distance between the nuclei is large 
enough that no purely hadronic interactions can occur. This roughly 
means impact parameters larger than the sum of the nuclear radii. 
The interaction is then instead mediated by the electromagnetic field. 
For a recent review of ultra-peripheral collisions, see \cite{ARNPS}.

\section{Ultra-peripheral collisions at RHIC}

The Relativistic Heavy-Ion Collider (RHIC) at Brookhaven National Laboratory 
began operating in 
the year 2000. This meant an increase in the maximum center of mass energies 
for heavy-ion collisions by more than an order of magnitude compared with the 
earlier fixed-target experiments. 

At very high collision energies, the electromagnetic 
field surrounding a nucleus contains photons energetic enough to 
produce new particles in ultra-peripheral collisions. This can happen in a 
purely electromagnetic process through a two-photon interactions or in an 
interaction between a photon from one of the nuclei and the other (``target'') nucleus. 
The photon spectrum for a minimum impact parameter, $b_{min}$, 
extends to $\sim \gamma / b_{min}$, which corresponds 
to about 300 GeV in the rest frame of the target nucleus in a gold on 
gold collision at RHIC. These photon energies are thus far above the 
threshold for particle production. The coherent contribution from the 
$Z$ protons in the nucleus, furthermore, enhances the number of equivalent 
photons by a factor $Z^2$. 

The high photon energies and fluxes lead to large cross sections for 
several photon-induced reactions; some have 
cross sections much larger than the total hadronic cross section and 
are major sources of beam-loss at heavy-ion colliders\cite{Baltz}. 
For example, the cross section for breaking up 
one of the nuclei in an Au+Au collision at RHIC through a photonuclear 
interaction is 95~b. 
The cross section for exchanging two photons and thereby simultaneously 
breaking up both nuclei in the same event is also large, about 4~b. 
The dominating fragmentation mechanism is excitation to a Giant Dipole 
Resonance followed by emission of one or a few neutrons. 

The mutual Coulomb dissociation has been studied at RHIC by detecting 
the forward going neutrons in Zero-Degree Calorimeters\cite{Chiu}. 
These are located 18~m 
downstream from the interaction point and have an angular acceptance 
of $\theta <$~2~mrad with respect to the beam axis. 
The relative contribution of the photon-induced fragmentation to the total 
cross section was found to be in good agreement with calculations based on 
the method of equivalent photons combined with measured $\gamma$+Au 
cross sections. The calculations also reproduced the neutron multiplicity 
distribution for photon-induced events reasonably well. 

Another ultra-peripheral process with very high cross section is two-photon 
production of electron-positron pairs. Of particular interest is the sub-class 
of events where the produced electron binds to one of the beam nuclei. 
The captured electron changes the charge and thus the rigidity of the ion, 
leading to a different deflection by the guiding magnets in the accelerator 
ring and eventual loss. Under certain conditions, the ion with an attached 
electron will 
hit the wall of the beam-pipe enclosure at a well-defined spot down stream 
of the interaction point. At the Large Hadron Collider at CERN, 
because of the high beam flux and energy, this has the potential to 
heat and quench the superconducting magnets near this area. The phenomenon 
was recently observed for the first time at RHIC with Cu--beams\cite{BFPP}. 
The location of the point of incidence ($\approx$140~m downstream from the 
interaction point) and the multiplicity of secondary 
particles resulting from the interaction of the 100 A GeV Cu beam with 
the beam-pipe and the surrounding magnets were found to be in good agreement 
with theoretical calculations, although the experimental uncertainties 
were large. 

Particle production in ultra-peripheral collisions has been studied by 
both of the two large experiments at RHIC, STAR and PHENIX. Some of these  
results will be discussed in the following two sections.

\section{Results from STAR} 

The first results on particle production in ultra-peripheral collisions at
RHIC were studies of coherent production of $\rho^0$ mesons in Au+Au 
interaction by the STAR collaboration\cite{STARRho}. 
The cross section to produce a 
$\rho^0$ in an Au+Au collision at RHIC is about 10\% of the total
inelastic, hadronic cross section.

STAR has also published final results on two-photon production of free 
$e^+ e^-$--pairs\cite{STARee} and 
preliminary results on photo-production of $\rho^0$ in 
d+Au collisions\cite{STARdAu} and coherent production of 
four pions in Au+Au collisions\cite{STAR4pi}.  

In d+Au collisions more than 90\% of the photo-produced $\rho^0$ mesons come 
from events where the gold nucleus emitted the photon. 
The interactions can leave the deuteron intact 
$\gamma + d \rightarrow \rho^0 +d$ or lead to break-up 
$\gamma + d \rightarrow \rho^0 +n+p$. Two triggers were implemented to study 
the two cases. Both were based on triggering on low multiplicity combined with 
a ``topology'' cut to reject cosmic rays. The multiplicity was measured in the 
STAR Central Trigger Barrel, which consists of 240 scintillators covering the 
full azimuth in the pseudo-rapidity range $|\eta| <$~1. To trigger on interactions 
where the deuteron breaks up, it was in addition required that the forward going 
neutron should be detected in the Zero Degree Calorimeter. 
Examples of the $\pi^+ \pi^-$ invariant mass distributions for the two 
samples are shown in Fig.~1. 

\begin{figure}[htbp]
\begin{center}
\mbox{\hspace*{0.25cm} \epsfig{file=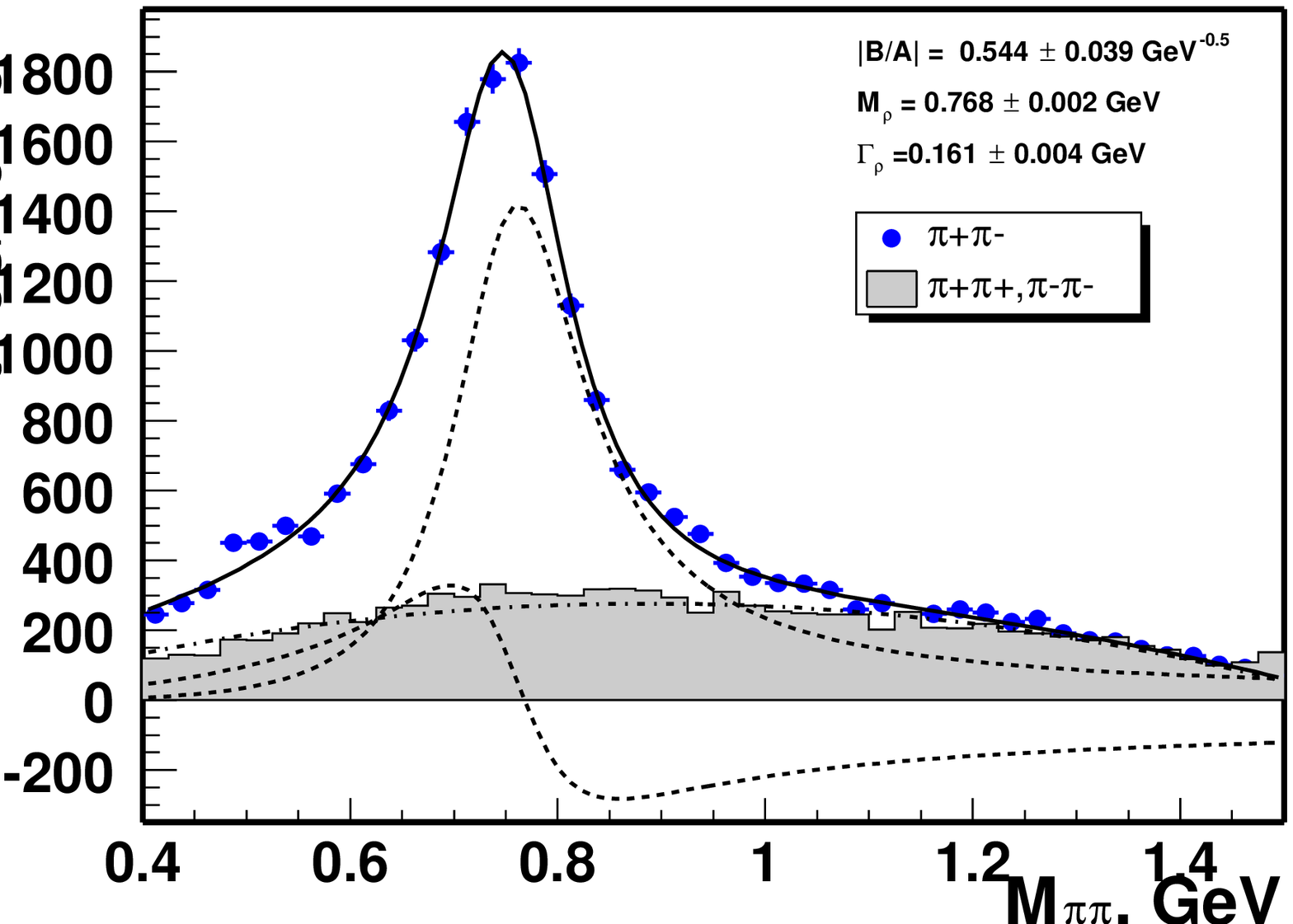,height=55mm,width=73mm} \hspace*{0.15cm} 
\epsfig{file=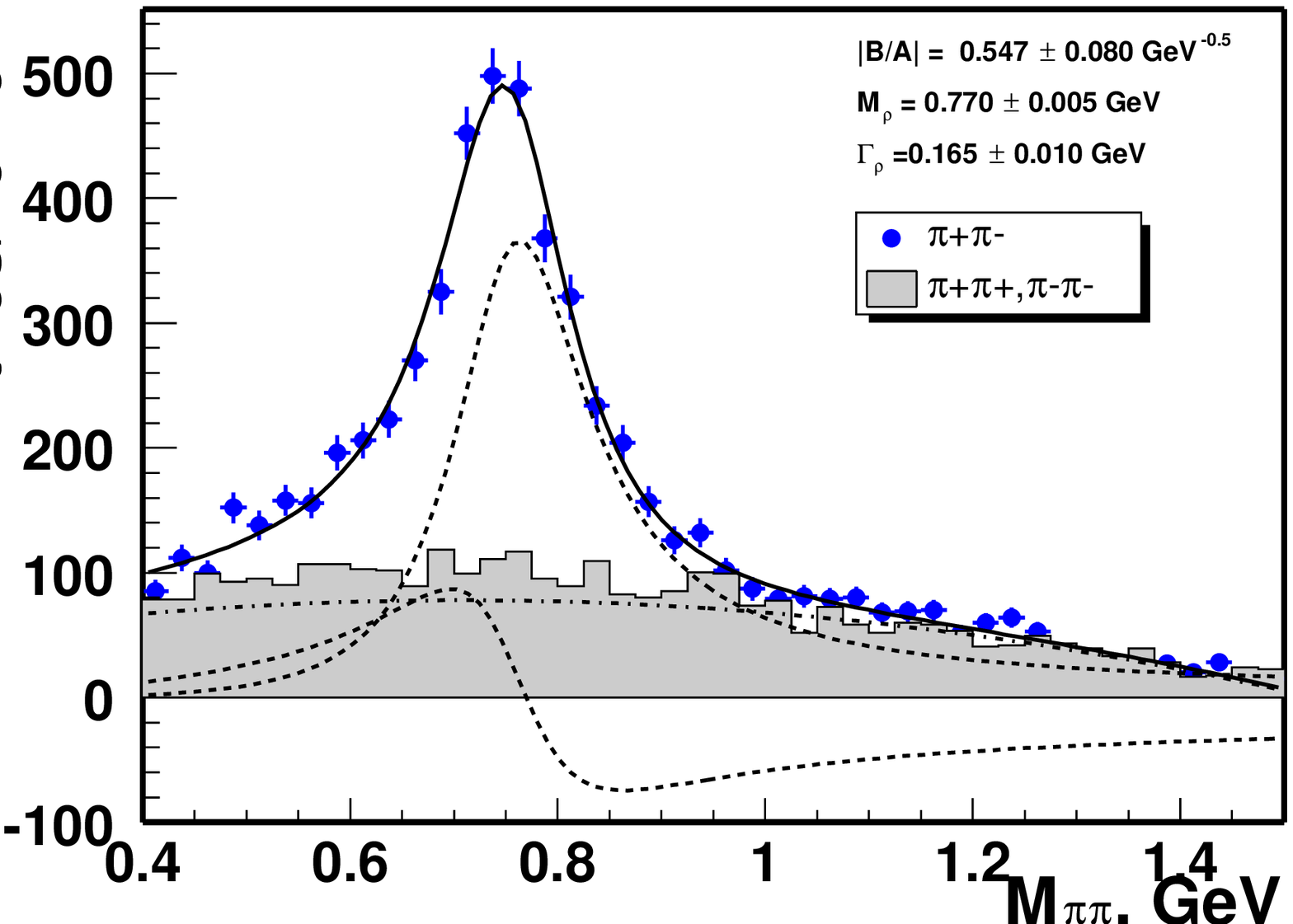,height=55mm,width=73mm}}
\caption{Invariant mass distributions for photoproduction of 
$\pi^+ \pi^-$--pairs in d+Au interactions at RHIC. The left (right) figure 
is for reactions where the deuteron remains intact (breaks up). 
From~[7].} 
\end{center}
\label{STARFigure}
\end{figure}

The invariant mass distribution is well described by the sum of the 
amplitudes for a resonant $\rho^0$ term and a non-resonant (S{\"o}ding) term: 
\begin{equation}
\frac{d \sigma}{dm_{\pi \pi}} = \left| 
A \frac{ \sqrt{m_{\pi \pi} m_{\rho} \Gamma_{\rho}} }{m_{\pi \pi}^2 
- m_{\rho}^2 +i m_{\rho} \Gamma_{\rho} } 
+ B \right|^2 + f_p
\end{equation}
Here, $\Gamma_{\rho} = \Gamma_0 \cdot (m_{\rho}/m_{\pi \pi}) \cdot 
[(m_{\pi \pi}^2 - 4 m_{\pi}^2)/(m_{\rho}^2 - 4 m_{\pi}^2)]^{3/2}$ is the 
momentum dependent width,  $\Gamma_0$ the natural width, and 
$f_p$ is a second order polynomial describing the background (estimated 
from the like-sign yield).  The mass and width of the $\rho^0$ are 
consistent with the Particle Data Group values. The ratio $A/B$ is a measure of the 
relative resonant to non-resonant contribution. The values observed in 
d+Au interactions (0.544 and 0.547 GeV$^{-1/2}$) are consistent, within 
errors, with the value obtained in Au+Au interactions, 
$0.81 \pm 0.08 \pm 0.20$~GeV$^{-1/2}$\cite{STARRho}.

\section{Results from PHENIX} 

The PHENIX experiment has studied the production of high-mass 
($m_{inv} >$~1.6~GeV) $e^+ e^-$-pairs and $J / \Psi$ 
mesons in ultra-peripheral Au+Au collisions\cite{DdE}.
A trigger was implemented for events where at least one of the nuclei 
break up through Coulomb dissociation. The trigger required a  
cluster with energy deposit $E > 0.8$~GeV in the 
Electromagnetic Calorimeter in coincidence with a signal in 
one of the Zero Degree Calorimeters and in anti-coincidence with a signal 
in the Beam-Beam Counters. The latter are Cherenkov counters covering 
$3.0 < |\eta| < 3.9$ and the absence of a signal corresponds to a rapidity 
gap on each side of the produced particles. 

In the offline analysis, the electron and positron were identified by 
the Ring Imaging Cherenkov Counters and Electromagnetic Calorimeters. 
These detectors are part of the PHENIX mid-rapidity tracking arms and  
cover $2 \times 90^o$ in azimuth and $|\eta|<0.35$ in pseudo-rapidity.  
Signal events were defined as those events with exactly one reconstructed 
$e^+$ and one reconstructed $e^-$ in opposite tracking arms. Events 
with a like-sign pair were used to estimate the amount of background. 

The transverse momentum and invariant mass 
distributions with the background from like-sign pairs subtracted are 
shown in Fig.~2. The transverse momentum is here the absolute value of the 
vector sum of the transverse momenta of the the $e^+$ and $e^-$. 

\begin{figure}[htbp]
\begin{center}
\mbox{\epsfig{file=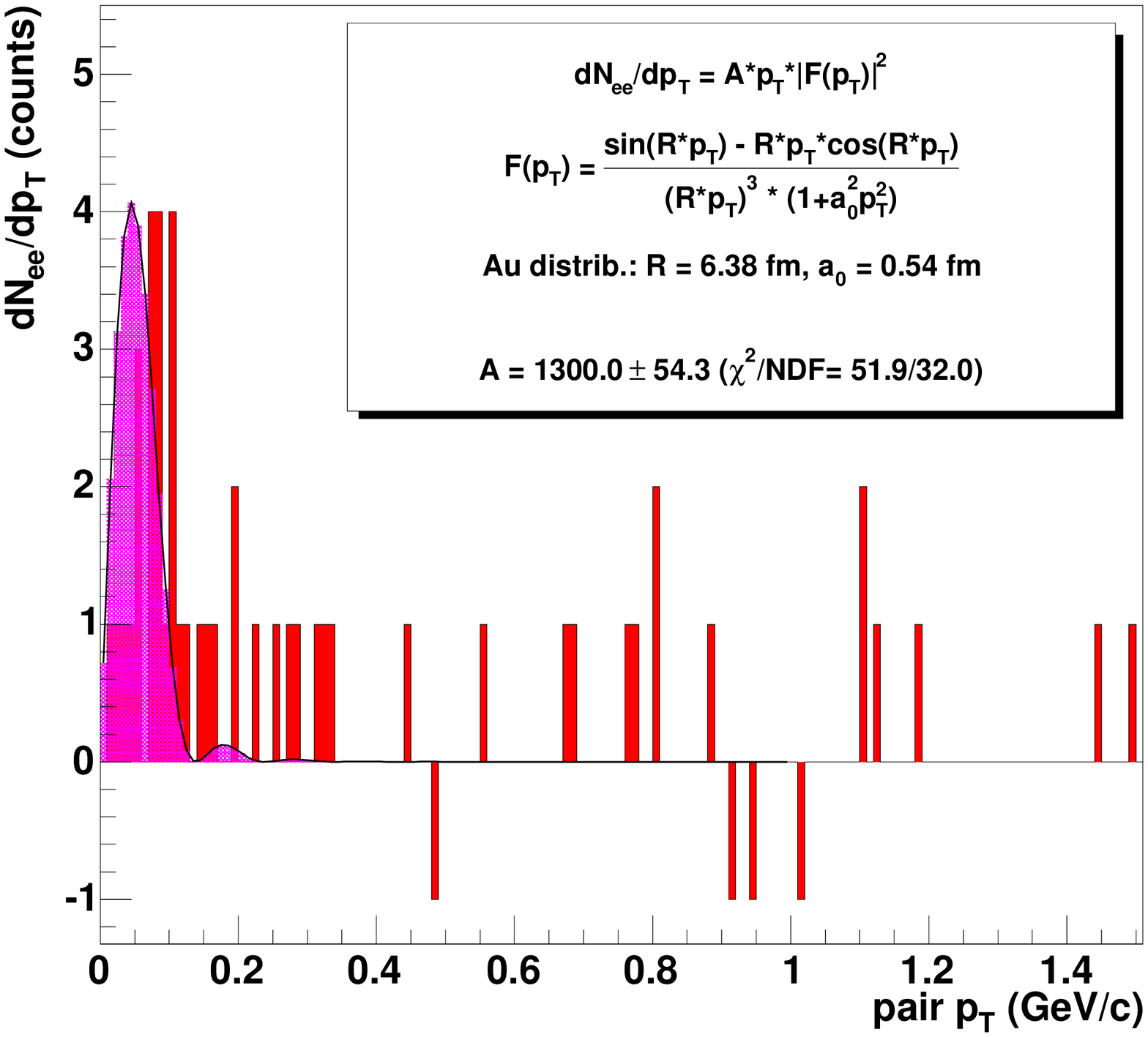,height=50mm,width=70mm}\quad\quad
\epsfig{file=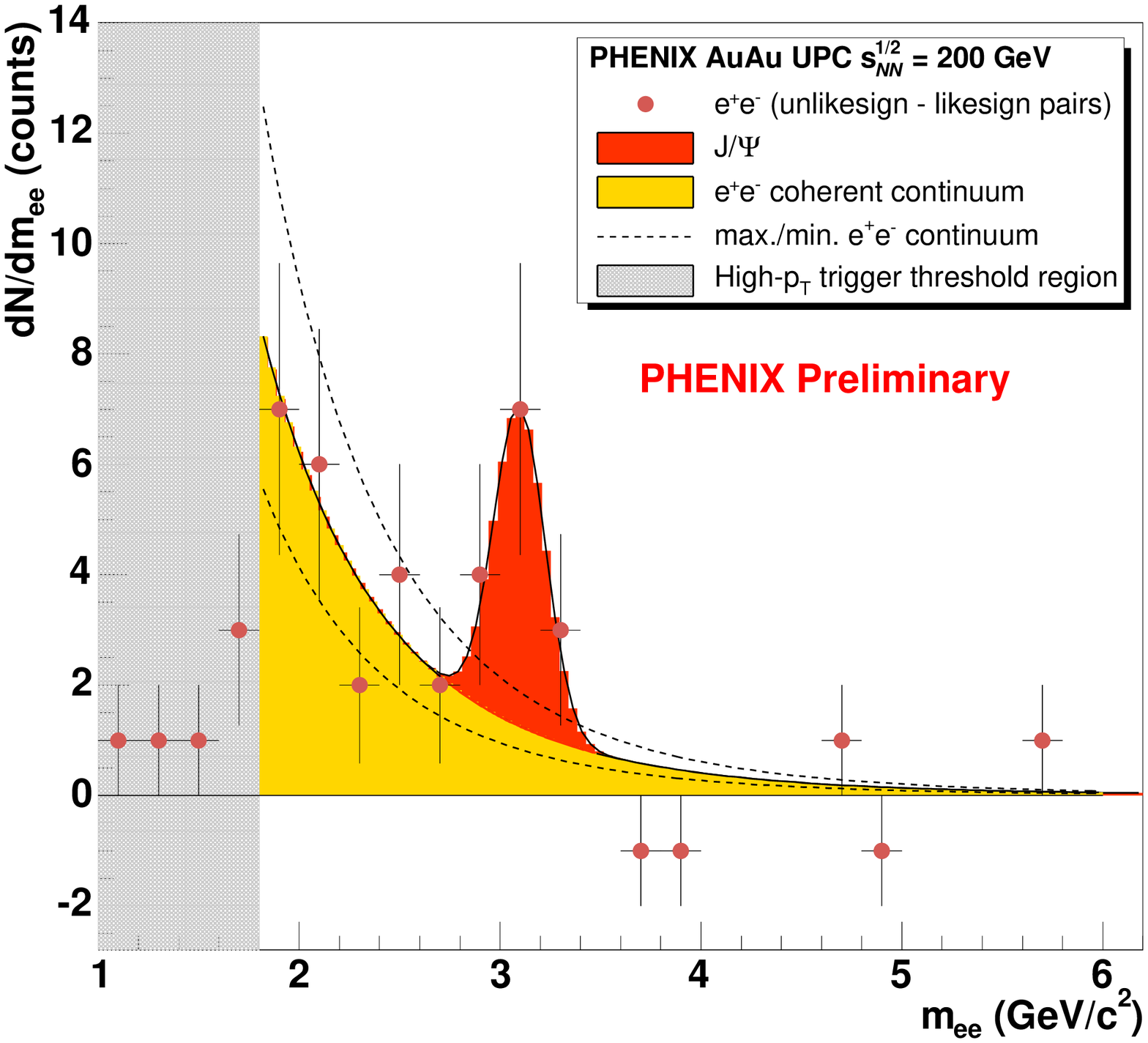,height=50mm,width=70mm}}
\caption{Transverse momentum (left) and invariant mass (right) 
distributions for 
$e^+ e^-$-pairs produced in ultra-peripheral Au+Au collisions. The 
background estimated from events with like-sign pairs have been subtracted 
(hence the negative entries in some bins). The curve in the left figure 
corresponds to the nuclear form factor. The solid curve in the right 
figure is a fit to the sum of a continuum and $J/\Psi$ 
distribution. The two additional dashed curves indicate the maximum and 
minimum continuum contributions. From~[9].}
\end{center}
\label{PHENIXFigure}
\end{figure}

The results show that exclusive production of high-mass $e^+ e^-$-pairs 
in Au+Au interactions at RHIC can be understood as a continuum contribution 
from two-photon production $\gamma + \gamma \rightarrow e^+ e^-$ and 
a contribution from photoproduction of $J/\Psi$s decaying into 
$e^+ e^-$-pairs. 

The cross section to produce a $J/\Psi$ in a coherent interaction is 
much larger than the cross section to produce a continuum pair in the corresponding  
mass range, but the branching ratio for $J / \Psi \rightarrow e^+ e^-$ 
(5.94 \%) and the broadening of the $J/\Psi$ peak because of the limited 
experimental resolution make the rates comparable in the relevant  
mass range\cite{JNINPC04}. 

The transverse momentum distribution shows a clear peak at very low 
transverse momenta, which is expected for coherent events where the 
momentum transfer is restricted by the nuclear form factor. 
The signal events with $p_T >$~100~MeV/c are believed to come from 
quasi-elastic $J/\Psi$ photoproduction, 
$\gamma + nucleon \rightarrow J/\Psi + nucleon$\cite{strikman}.

The total net number of $e^+ e^-$-pairs in the sample 
(integrated luminosity $120 \pm 10$~$\mu$b$^{-1}$) is about 40, 
of which $\approx$10 are estimated to come from decay of $J/\Psi$s. 
The observed rates are in reasonable agreement with expectations 
for $\gamma \gamma \rightarrow e^+ + e^-$ and photoproduction of 
$J/\Psi$\cite{JNINPC04}.

\section{Ultra-peripheral proton-proton collisions}

Electromagnetic interactions can of course also be studied with beams of 
protons or anti-protons, but there is then no enhancement ($\propto Z^2$) in the  
photon flux. Although evidence for electromagnetic particle production 
in $pp$ collisions were observed at the ISR almost 30 years ago\cite{ISR}, 
electromagnetic interactions in $pp$ collisions have so far attracted 
relatively little attention, much less than particle production in 
doubly diffractive interactions, for example. The CDF Collaboration 
has, however, recently published its first paper on two-photon production
of $e^+ e^-$--pairs with $m_{inv} >$~10~GeV in $p \overline{p}$ collisions
at the Tevatron\cite{CDFgammagamma}. 

The choice of minimum invariant mass is unfortunate, since it falls right in 
the range of the $\Upsilon (1S)$, $\Upsilon (2S)$, and $\Upsilon (3S)$ 
vector mesons, and, as will be shown, these vector mesons are expected to 
give a significant contribution to the exclusive production of 
$e^+ e^-$--pairs through the decay $\Upsilon \rightarrow e^+ e^-$. 

\begin{figure}[htbp]
\begin{center}
\epsfig{file=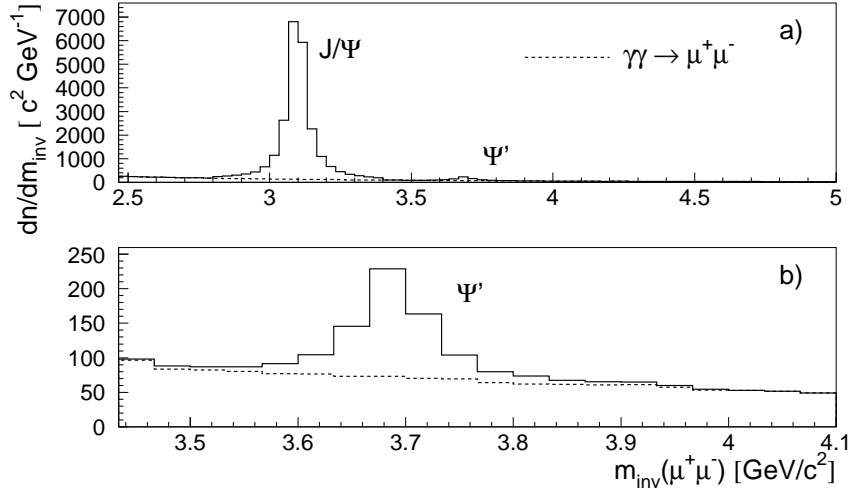,height=78mm}
\caption{Calculated invariant mass distributions for electromagnetic production 
of $\mu^+ \mu^-$--pairs around mid-rapidity ($|\eta| <$~2) in 
$\sqrt{s}$ = 1.96~TeV $p \overline{p}$ collisions. Figure b) shows the same 
distribution as a) but restricted to a narrow intervall around the $\Psi'$ mass. 
The widths of the $J/\Psi$ and $\Psi'$ have been set to roughly correspond to 
typical experimental widths.}
\end{center}
\end{figure}

CDF is also analyzing the exclusive production of $\mu^+ \mu^-$--pairs, 
$p \overline{p} \rightarrow p \overline{p} + \mu^+ \mu^-$, at lower invariant 
masses\cite{AH}. The two main contributions to these events are, as with 
heavy-ion beams, $\gamma \gamma \rightarrow \mu^+ + \mu^-$ and 
$\gamma + $Pomeron~$\rightarrow J/\Psi$ or $\Psi'$, followed by decay 
of the vector meson to a dilepton pair. There could also be a contribution 
from the elusive Odderon through the reaction 
Odderon+Pomeron $\rightarrow V \rightarrow \mu^+ + \mu^-$ \cite{Lech,odderon}.
The size of this contribution is not very well known; 
it is possible that it is comparable to the other two processes. 

There could also be a background to the $J / \Psi$ production from 
Pomeron+Pomeron interactions producing a $\chi_c$ via the decay 
$\chi_c \rightarrow J/\Psi + \gamma$ where the photon escapes 
detection. This background is a problem only for the $J / \Psi$ and 
not for the $\Psi'$. 

Figure~3 shows the calculated yields for continuum 
$\mu^+ \mu^-$--production and photoproduction of $J/\Psi$ and $\Psi'$ 
followed by decay to $\mu^+ \mu^-$--pairs for $p \overline{p}$ collisions
at the Tevatron ($\sqrt{s}$ = 1.96~TeV). It is required that both muons 
are within $|\eta| <$~2 to simulate the typical experimental acceptance 
at high-energy colliders. The vector meson production is  
calculated as in \cite{VMpp}, but with a larger cut-off for the minimum 
impact parameter ($b >$~1.4~fm rather than 0.7~fm). This is believed to 
better reproduce the condition of no accompanying hadronic interactions. 
The input $\gamma + p \rightarrow p + \Psi'$ cross section is taken 
from\cite{H1}. The continuum $\gamma + \gamma \rightarrow \mu^+ \mu^-$ 
is calculated under the same conditions as the vector mesons. 
The contribution from two-photon interactions is much smaller in 
$pp$ or $p \overline{p}$ collisions compared with in heavy-ion collisions 
(cf. Fig.~2). The calculated cross sections are given in 
Table~1. 

\begin{table}[bth]
\caption{Cross sections for photoproduction of vector mesons and 
$\mu^+ \mu^-$--pairs in $p \overline{p}$ collisions at the Tevatron ($\sqrt{s}$ = 1.96~TeV). 
The rightmost column shows the cross section multiplied with the branching ratio for 
decay into $\mu^+ \mu^-$.}
\renewcommand{\tabcolsep}{1.15pc} 
\renewcommand{\arraystretch}{1.2} 
\begin{center}
\begin{tabular}{lcc}
\hline
               & $\sigma$ $[$nb$]$  & $\sigma \cdot Br(\mu^+ \mu^-)$ $[$nb$]$ \\ \hline
$p+\overline{p} \rightarrow p+\overline{p} + J/\Psi$ & 15 & 0.87 \\ 
$p+\overline{p} \rightarrow p+\overline{p} + \Psi'$ & 2.4 & 0.018 \\ 
$p+\overline{p} \rightarrow p+\overline{p} + \mu^+ \mu^-$ ($m_{inv} >$~1.5 GeV) & 2.4 & 2.4 \\ 
\hline
\end{tabular}
\end{center}
\end{table}

\section{Summary}

The fact that particles are produced in ultra-peripheral collisions and 
that the experiments that were designed to study central collisions can 
detect them have been shown by the STAR and PHENIX experiments at RHIC. 
Hopefully, the increase in the cross sections with energy and the extended 
trigger capabilities of future experiments will lead to an increased 
interest for ultra-peripheral collisions in proton-proton and 
heavy-ion collisions at the Large Hadron Collider.

\begin{footnotesize}
\bibliographystyle{nystrand} 
{\raggedright
\bibliography{nystrand}
}
\end{footnotesize}
\end{document}